\documentclass[10pt]{article}
\usepackage{amsmath, natbib}
\usepackage{graphicx}
\usepackage{authblk}
\usepackage[top=4cm, bottom=4cm, left=5cm, right=5cm]{geometry}
\usepackage{fancyhdr}
\renewenvironment{abstract}{%
\hfill\begin{minipage}{0.95\textwidth}
\rule{\textwidth}{1pt}}
{\par\noindent\rule{\textwidth}{1pt}\end{minipage}}
\makeatletter
\renewcommand\@maketitle{%
\hfill
\begin{minipage}{0.95\textwidth}
\vskip 2em
\let\footnote\thanks 
{\LARGE\bf \@title \par }
\vskip 1.5em
{\large \@author \par}
\end{minipage}
\vskip 1em \par
}
\makeatother
\begin{document}
%
\title{Reclassifying symbiotic stars with 2MASS and WISE: An atlas of spectral energy distribution}
\author[1]{Akras Stavros}
\author[2]{Guzman-Ramirez Lizette}
\author[2]{Leal-Ferreira Marcelo}
\author[3]{Ramos-Larios Gerardo}
\affil[1]{Observat\'{o}rio Nacional/MCTIC, Rio de Janeiro, Brazil}
\affil[2]{Leiden Observatory, Leiden University, Netherlands}
\affil[3]{Instituto de Astronom\'{i}a y Meteorolog\'{i}a, Mexico}

\maketitle

\begin{abstract}
We present a new updated catalogue of Galactic and extragalactic symbiotic stars (SySts). Since the last catalogue of SySts 
(Belczynski et al. 2000), the number of known SySts has significantly increased. Our new catalogue contains 
316 known and 82 candidates SySts. Of the confirmed Systs 252 are located in our Galaxy and 64 in nearby galaxies. This reflects an increase of 
$\sim$50\% in the population of Galactic SySts and $\sim$400\% in the population of extragalactic SySts. 
The spectral energy distribution (SED) 
of 334 (known and candidates) SySts have been constructed using the 2MASS and WISE data. These SEDs are used to 
provide a robust reclassification in scheme of S- (74\%), D- (15\%) and D'-types (2.5\%). The SEDs of S- and D-type peak between 
0.8 and 1.6~$\mu$m and between 1.6 and 4~$\mu$m, respectively, whereas those of D'-type exhibit a plateau profile. Moreover, 
we provide the first compilation of SySts that exhibit the OVI Raman-scattered line at 6830\AA. Our analysis shows that 
55\% of the Galactic SySts exhibit that line in their spectrum, whereas this percentage is different from galaxy to galaxy. 
\end{abstract}


\section{Introduction}

Symbiotic stars (SySts) are interacting, long-period, binary systems consisting of a cool red giant star that transfer matter
to a much hotter companion, usually a white dwarf but it is also possible to be a neutron star or black hole. The
atmosphere of the red giant or its wind are excited by the UV radiation from the white dwarf resulting in the formation
of a colourful nebula. Their spectrum consists of both absorption features due to the photosphere of the cool companion 
and a number of emission-lines from highly-excited ions due to the surrounding nebula.

SySts represent ideal objects for investigating and studying several astrophysical phenomena such as the formation
of aspherical circumstellar envelopes and high-velocity jets, dust forming regions, colliding winds, the interaction
of binary components and their evolution, mass transfer processes, accretion disks, soft and hard-X rays emission
(e.g. Mikolajewska 2012; Luna et al. 2013, Skopal \& Carikov\'{a} 2015; Mukai et al. 2016).
They have also been proposed as potential progenitors of type Ia supernova (SN Ia) due to the large
amount of masses that the white dwarf accretes from the cold companion resulting in exceeding the Chandrasekhar
mass (1.4 M$_{\odot}$) and exploding as a SN Ia (e.g. Di Stefano 2010; Dilday et al. 2012). 
Hence, the interest in SySts has been gradually increasing the last decates and many attempts are being made to 
discover new SySts either in our Galaxy or other galaxies in the Local Group.

Based on their near-IR 2MASS colours, SySts are divided into two main categories  near-infrared data: (i) those with a 
near-IR colour temperatures of $\sim$3000-5000 K, which attributed to the temperature of a G-, K- or M-type giant 
(stellar or S-type SySts), and (ii) those with a near-IR colour temperature around 700-1000 K, indicating a warm dusty 
circumstellar envelope (dusty or D-type SySts) (Webster \& Allen 1975). 

The identification of an object as SySts is made based on a number of widely used criteria: (i) the presence
of strong He~II and H~I emission lines as well as emission lines from high-excitation ions (ionization potential, I.P.$\geq$35~eV),
(ii) the presence of absorption features of TiO and VO associated with the photosphere of the cold companion and (iii) 
the presence of the OVI Raman-scattered line centred at 6830 \AA\ (e.g. Lee 2000; Belczy\'{n}ski et al. 2000; Mikolajewska et al. 2014).

More effort has been invested in developing a more general way of distinguishing SySts from other strong H$\alpha$ emitters 
(e.g. genuine PNe, H~II regions, WR stars, Be stars etc.). In the optical regime, Gutierrez-Moreno et al.(1995) proposed a 
diagnostic diagram between [O III] 4363/H$\gamma$ vs. [O III] 5007/H$\beta$ emission line ratios, which reflects on the different 
densities between PNe and SySts (see e.g. Clyne et al. 2015). Recently, Corradi et al. (2008) proposed a new diagnostic diagram 
based on the IPHAS r-H$\alpha$ vs. r-i colour indices.

SySts are also important X-ray sources. Based on their X-ray spectrum, they are divided into four types: 
(a) the supersoft X-ray sources with energies $\leq$0.4~kev probably emitted directly from the white dwarf ($\alpha$-type), 
(b) the objects that exhibit a peak at 0.8 kev in their X-ray spectrum and maximum energies up to 2.4 keV, likely
originate from a hot, shocked gas where the stellar winds collide ($\beta$-type), (c) the objects with a 
non-thermal emission and energies higher than 2.4 keV ($\gamma$-type) due to the accretion of mass onto the hot companion 
(white dwarf or neutron stars) and (d) those with very hard X-ray thermal emission and energies higher
than 2.4~keV likely originate from the inner regions of an accretion disk ($\delta$-type; Muerset et al. 1997; Luna et al. 2013).

\begin{figure}
\centering
\includegraphics[scale=0.30]{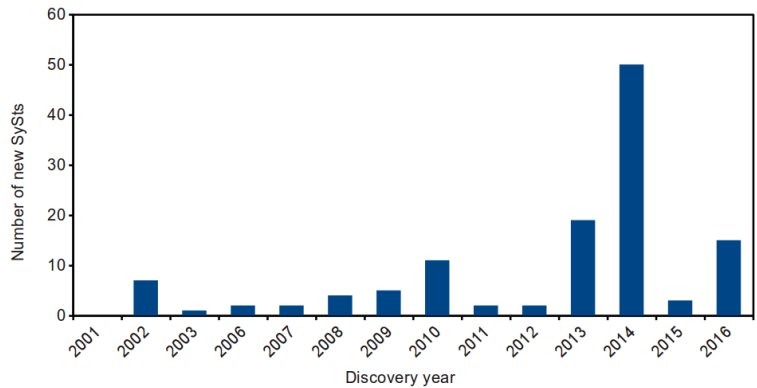}
\caption{The number of SySts discoveries per year between 2000 and 2016.}
\label{}
\end{figure}

\section{Sample selection}
The most complete and comprehensive compilation of SySts was published by Belczy\'{n}ski and collaborators 16
years ago (Belczy\'{n}ski et al. 2000). This catalogue includes all the known Galactic and extragalactic
SySts (188) as well as a number of 30 candidates SySts. The histogram in Figure 1, shows the number of new SySts 
discoveries per year the last 16 years. Of the 316 confirmed SySts, 252 are Galactic and 64 are extragalactic. This implies 
an increase of $\sim$50\% and $\sim$400\% in the population of Galactic and extragalactic SySts since 
the publication of Belczy\'{n}ski's catalogue (2000). 
Besides new discoveries, the total number of SySts still remains very low compared to the expected number of SySts in our 
Galaxy, which varies between 3$\times$10$^3$ (Allen 1984) and 4$\times$10$^5$ (Magrini et al. 2003).
Nevetheless, the number of SySts is expected to significantly increase over the next years due to the on-going surveys like 
VPHAS+ (Drew et al. 2014), J-PAS/J-PLUS (Benitez et al. 2014) and S-PLUS (Mendez de Oliveira et al. in prep.).

\section{Spectral energy distribution}
According to Ivison et al. (1995), spectral energy distribution of SySts peaks between 1 and 2$\mu$m for the S-type, 
5-15$\mu$m for the D-type and at longer wavelengths between 20 and 30$\mu$m for the D'-type SySts. Therefore, it is coherent 
to construct and study the SEDs of SySts using both the 2MASS and WISE data providing a more robust classification.

In Figure~2, we display two examples for each type of SySts (S-, D- and D'-type). The SED of S-type is dominated 
by the cool companion and those of D-type by the emission of dust. We performed a statistical analysis that shows 
the SED of the S-type peak between 0.8 and 1.6$\mu$m, D-type 1.6 and 5$\mu$m, whereas those of D'-type SySts show a plateau profile 
within the wavelength range covered by 2MASS and WISE. We also find a statistically significant number of SySts with 
a clear S-type profile plus an infrared excess at 11.6 and/or 22.1$\mu$m (see Fig.~2). We, thus, propose a fourth type of SySts, 
namely S+IR excess. A further study on this type of SySts is required in order to understand whether it consists a new type 
between the S- and D-type SySts. Figure~3 shows the percentages as well as the population of known and candidate SySts for each 
type based on our preliminary classification.

\begin{figure}
\centering
\includegraphics[scale=0.18]{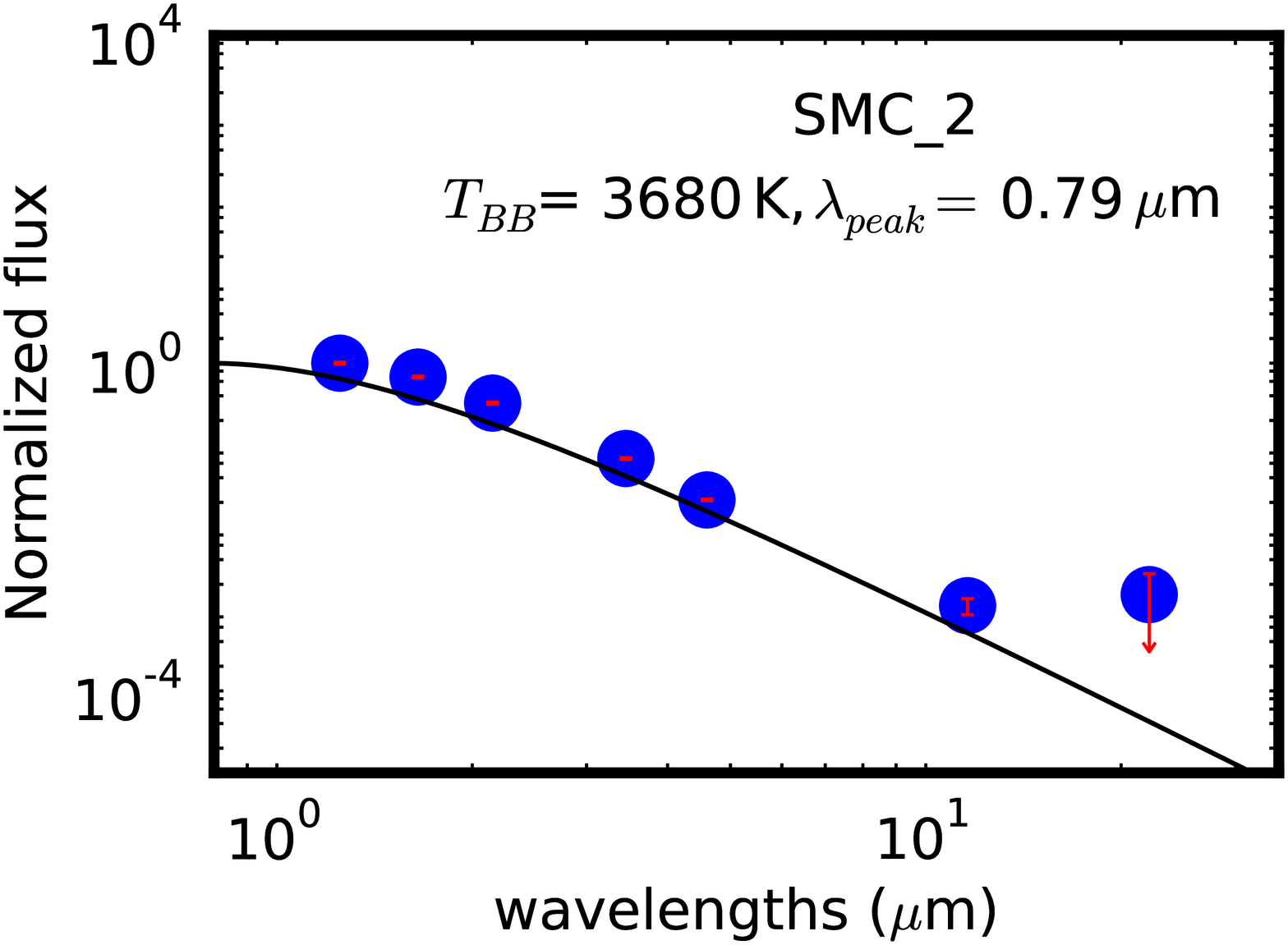}
\includegraphics[scale=0.18]{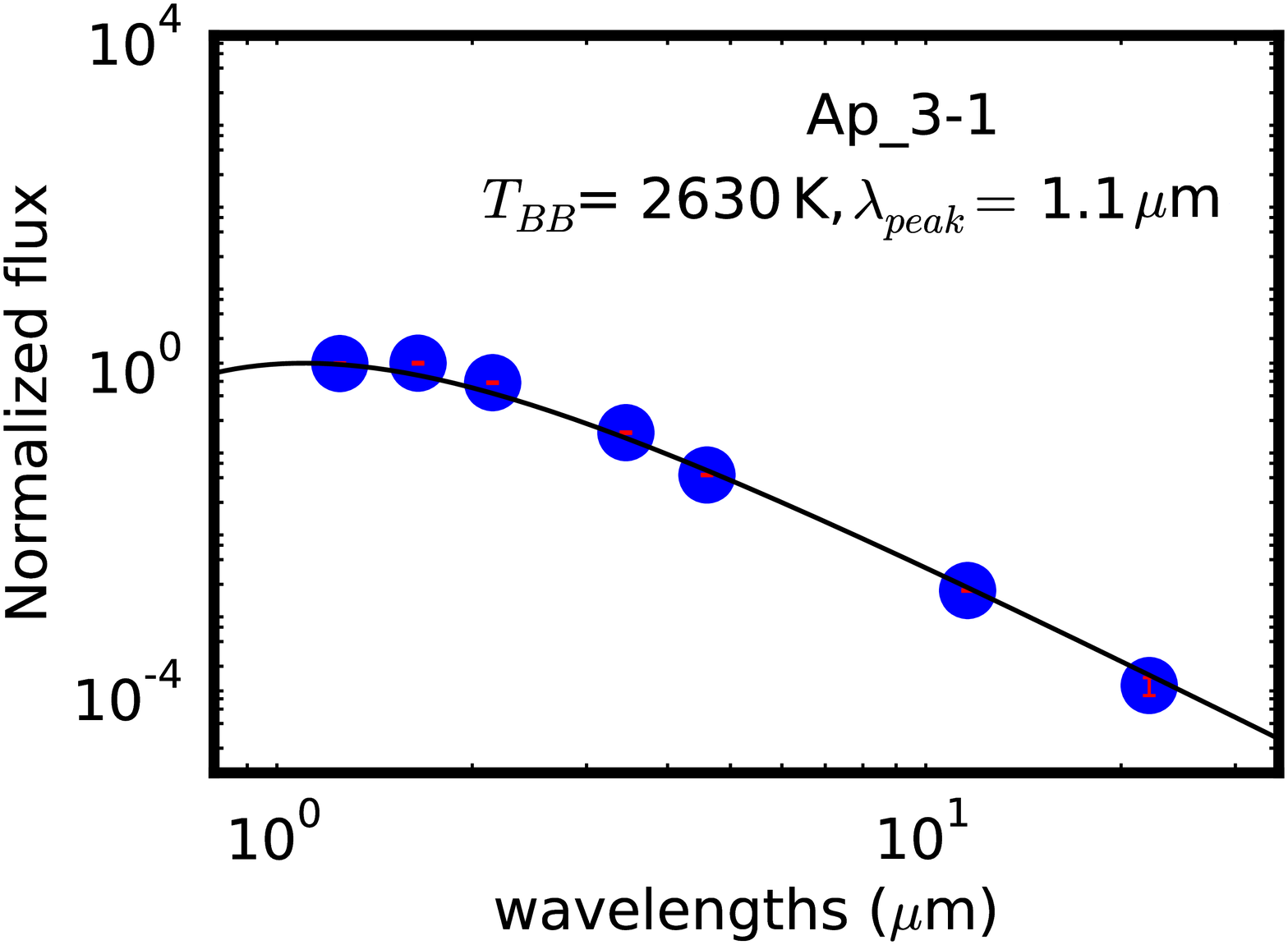}

\includegraphics[scale=0.18]{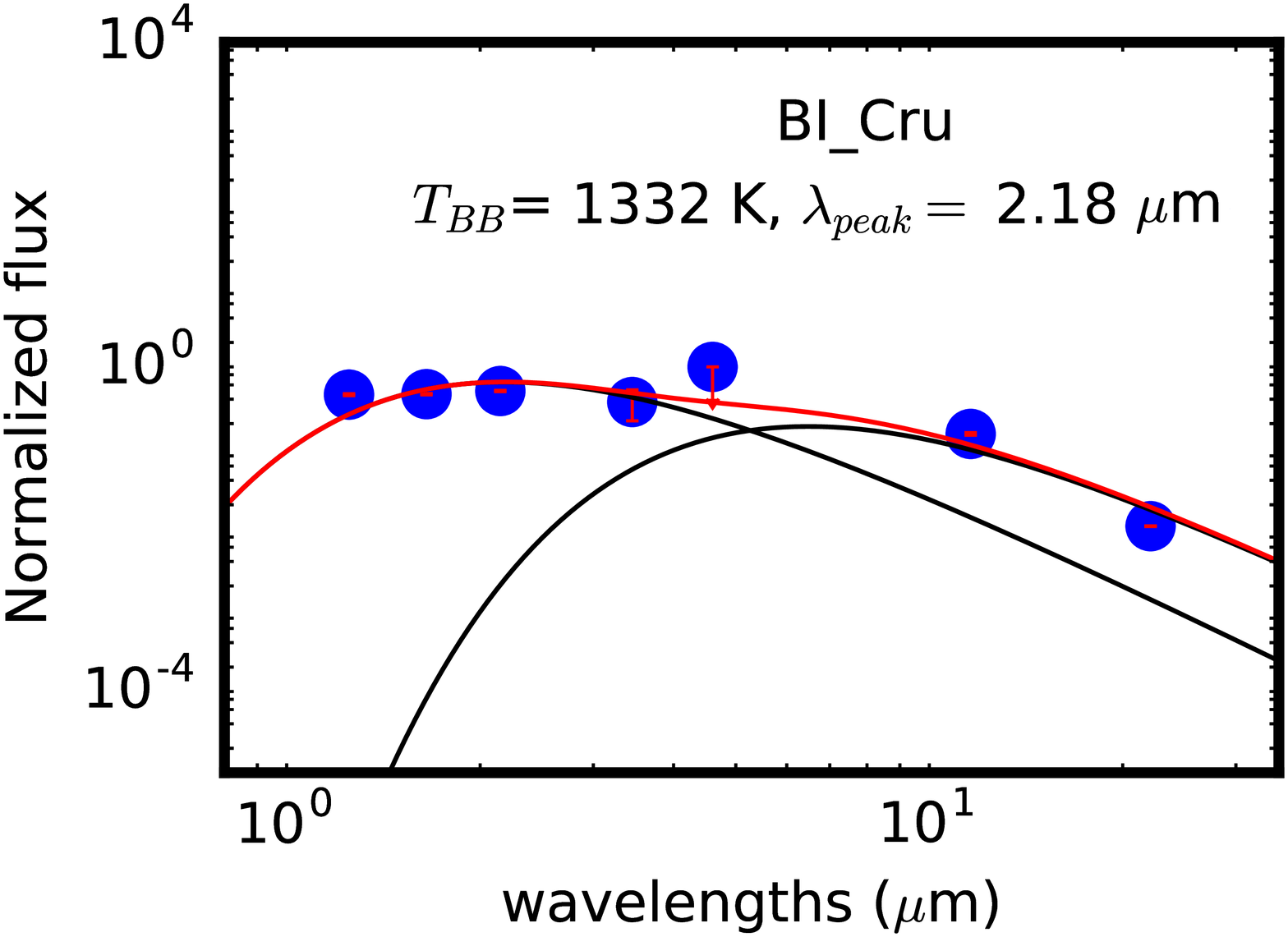}
\includegraphics[scale=0.18]{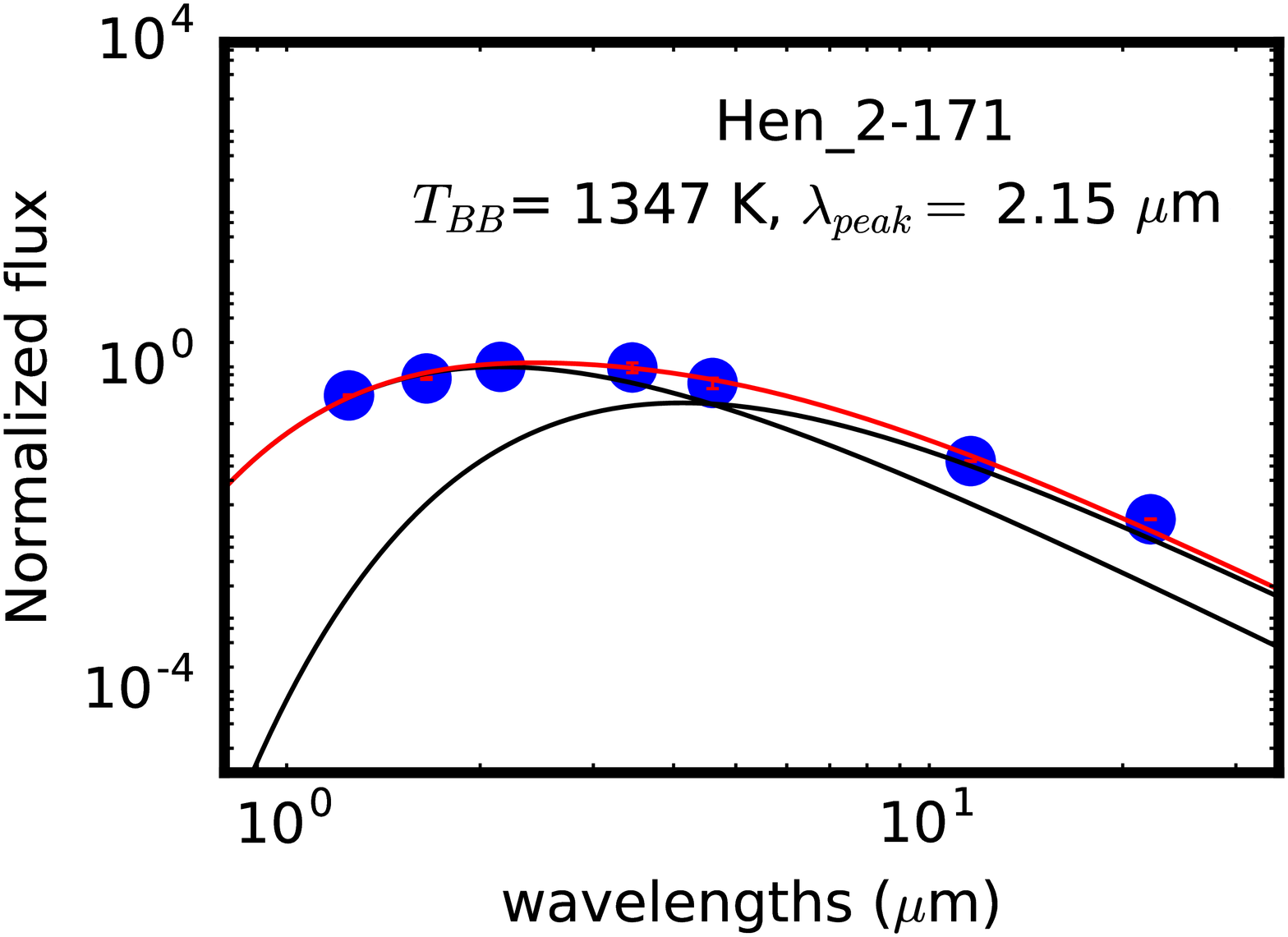}

\includegraphics[scale=0.18]{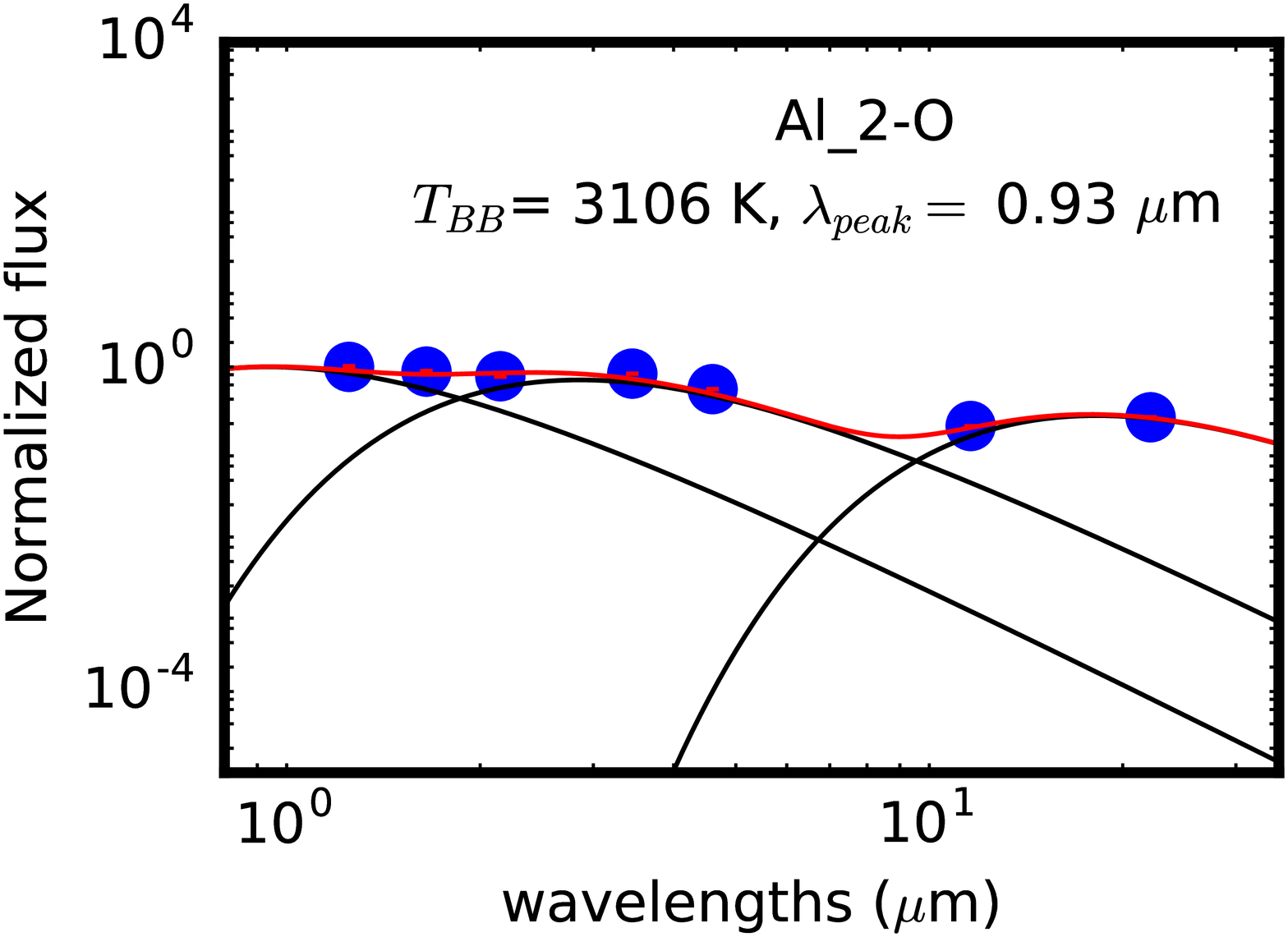}
\includegraphics[scale=0.18]{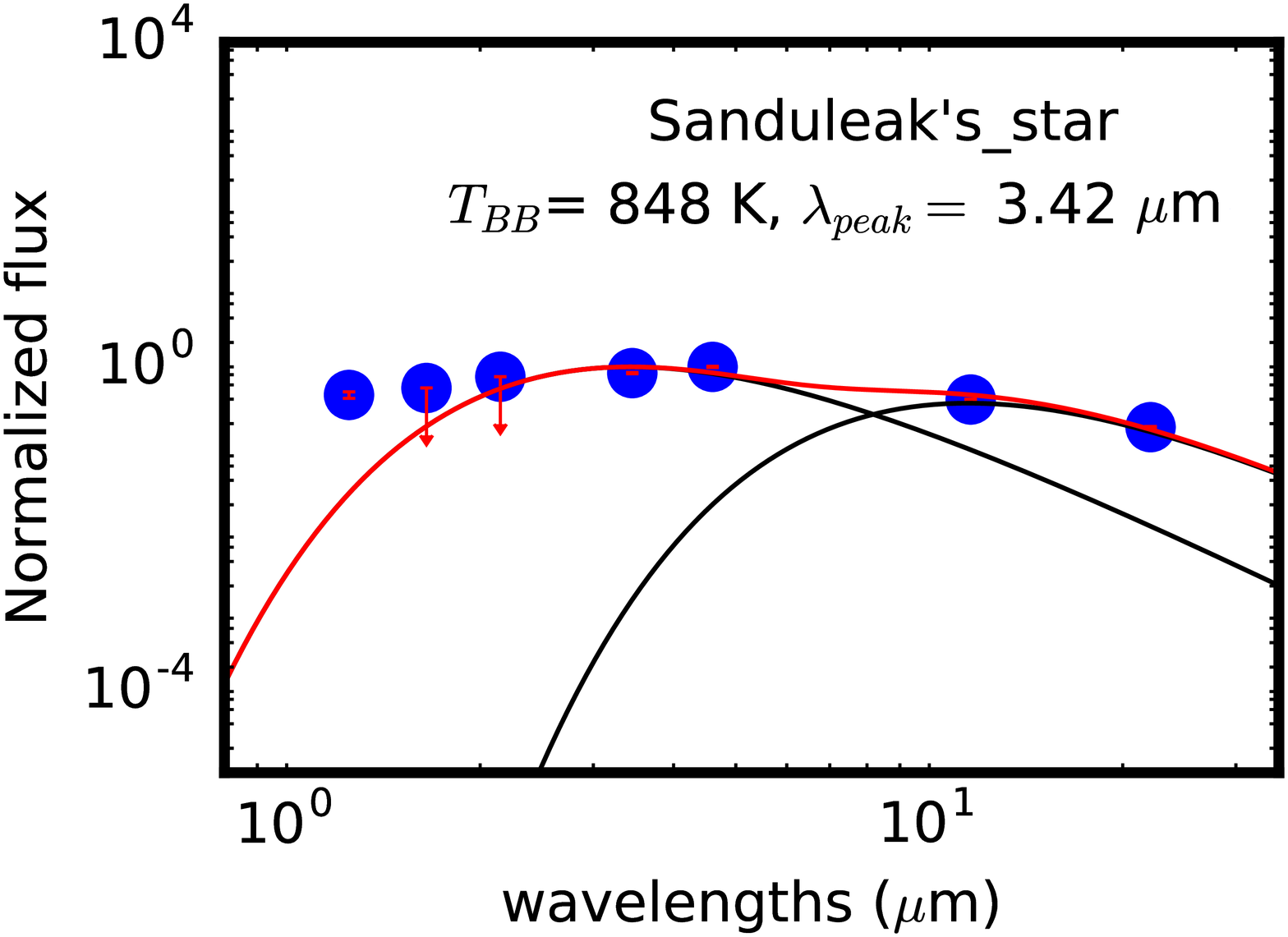}

\includegraphics[scale=0.18]{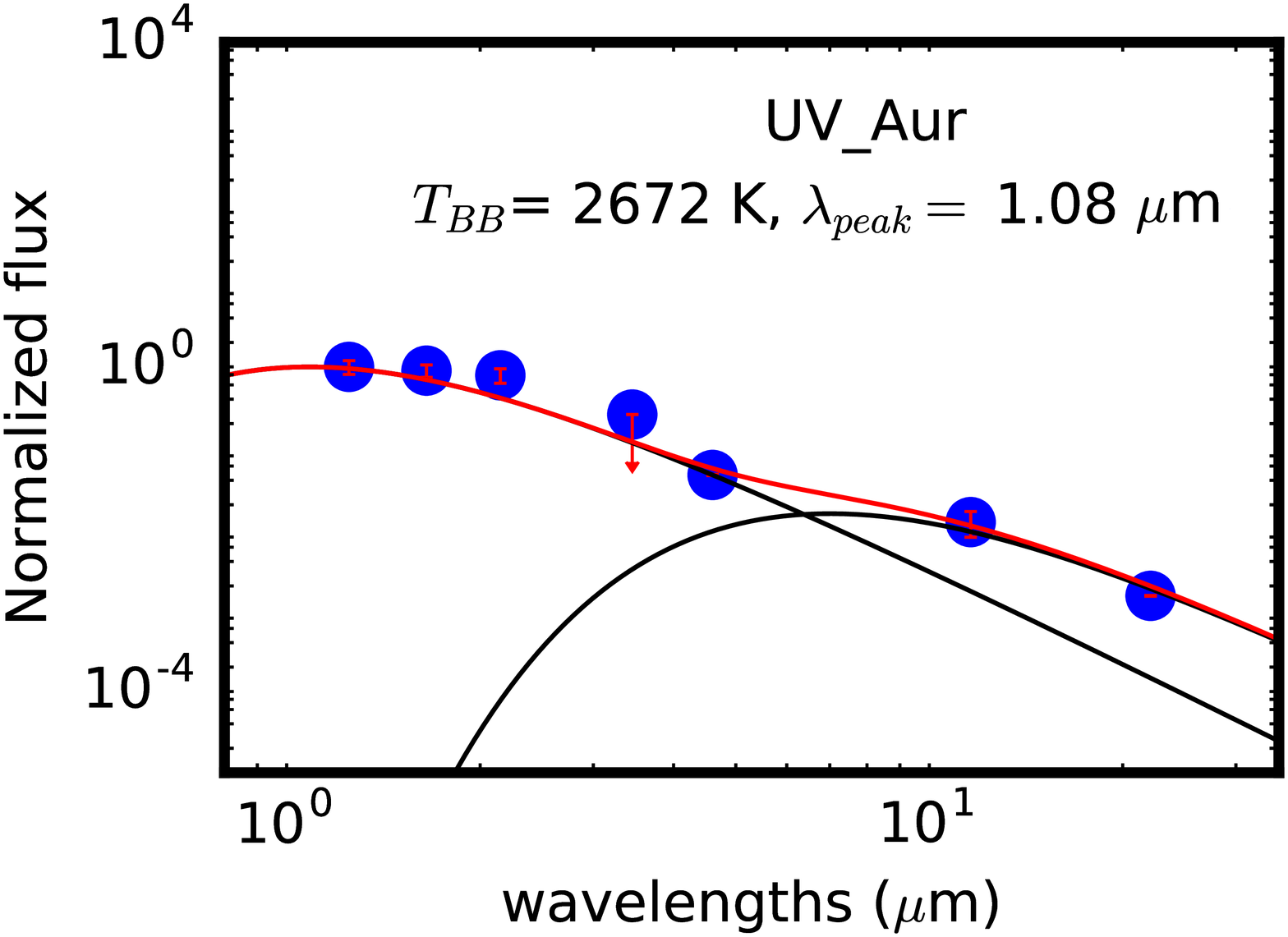}
\includegraphics[scale=0.18]{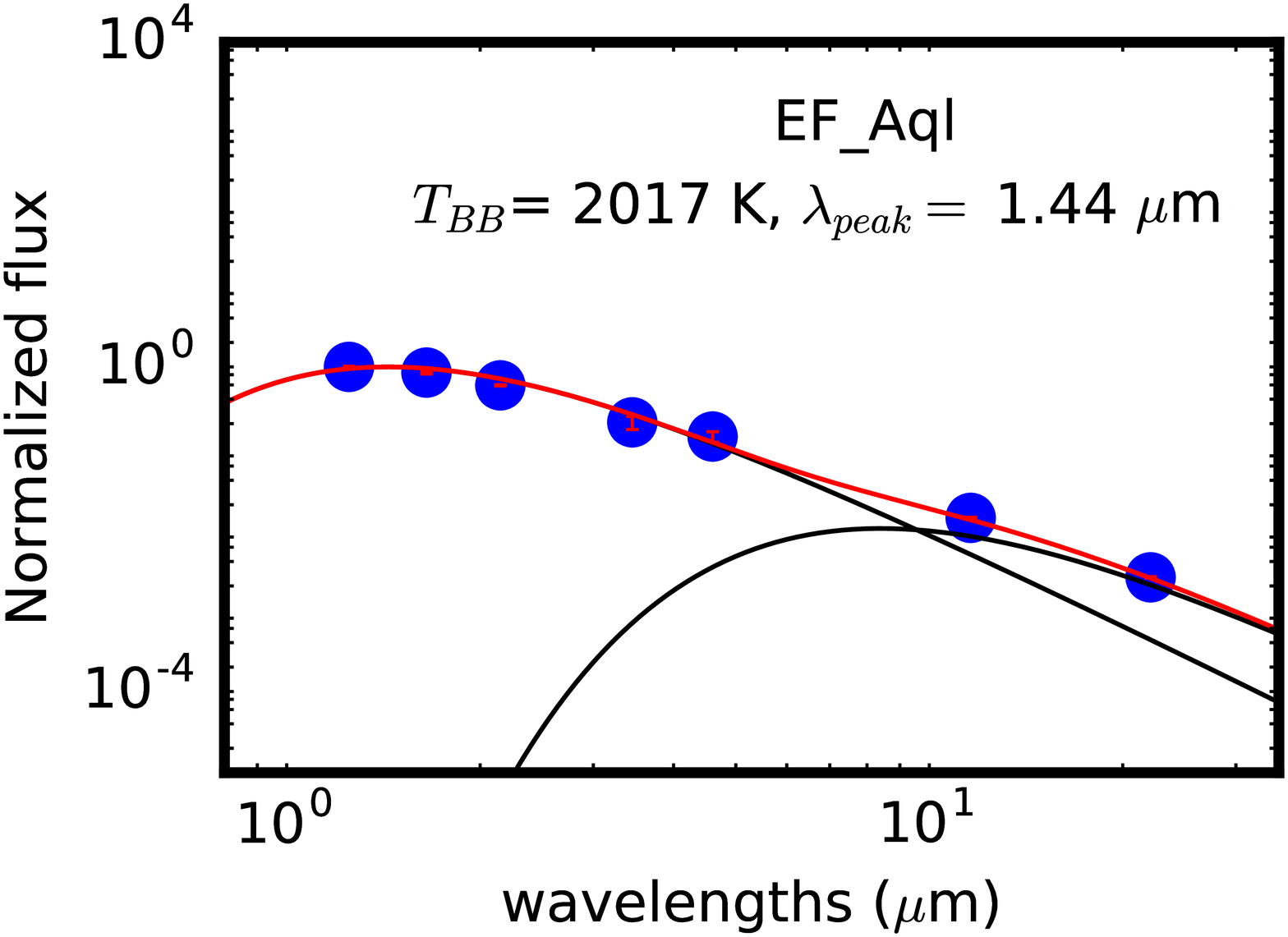}
\caption{Examples of SEDs for two S- (first line), D- (second line), D'-(third line) and S+IR-type (fourth line) SySts.}
\label{}
\end{figure}

\section{OVI Raman scattered line 6830\AA}
One of the criteria to classify an object as SySt is the detection of the line features centred at 6830\AA\ and 7088\AA\
(Belczynski et al. 2000). These two broad lines are interpreted as the result of Raman scattering of the ultraviolet OVI 
$\lambda\lambda$1032,1038 resonance lines by neutral hydrogen (Schmid 1989). Allen (1980) pointed out that 50\% or more of  
SySts exhibit these Raman lines even before their identifications. However, this analysis should be further explored since the sample 
contains only a few SySts.

Therefore, the first compilation of SySts that show the OVI Raman-scattered line $\lambda$6830 in their spectrum is presented 
in the current catalogue. We find that 55\% of the Galactic SySts (119 out of 218 with available optical spectra) show the OVI 
$\lambda$6830 line very close to what found by Allen (1980). Examining the detection of the OVI Raman line in different type, we 
find no difference between S- (58\%) and D-types (54\%), whereas only 8 out of 21 S+IR excess type (38\%) and one out 
of four D'-type (25\%) show the Raman line.

Given that the number of confirmed extragactic SySts has increased, we are able now to perform a similar analysis in four nearby 
galaxies. In particular, we find that all eight known SySts in the SMC show the Raman line (100\%), whereas only four out of seven in the LMC (57\%). As for the M31 and M33 galaxies, 16 out of 31 SySts in the M31 (51\%, Mikolajewska et al. 2014) and 5 out of 12 SySts in the M33 
(41\%, Mikolajewska et al. 2017) show the Raman line in their spectra. This implies that galaxies have different percentages probably 
due to different physical conditions such as the metallicity parameter. However, these sample contains very few SySts and this results has to be further explored in the future when the sample have a statistically significant number of SySts.  

Moreover, looking carefully at the spectra of the SySts in M31 and M33 (Mikolajewska et al. 2014,2017), 
we also find a clear trend between the OVI $\lambda$6830 Raman line and He~II $\lambda$4686. When the latter line is detected the former 
is also detected, the opposite is not true and the flux ration between the two line is F$_{O VI}$/F$_{He~4686}\sim$0.5.

\begin{figure}
\centering
\includegraphics[scale=0.35]{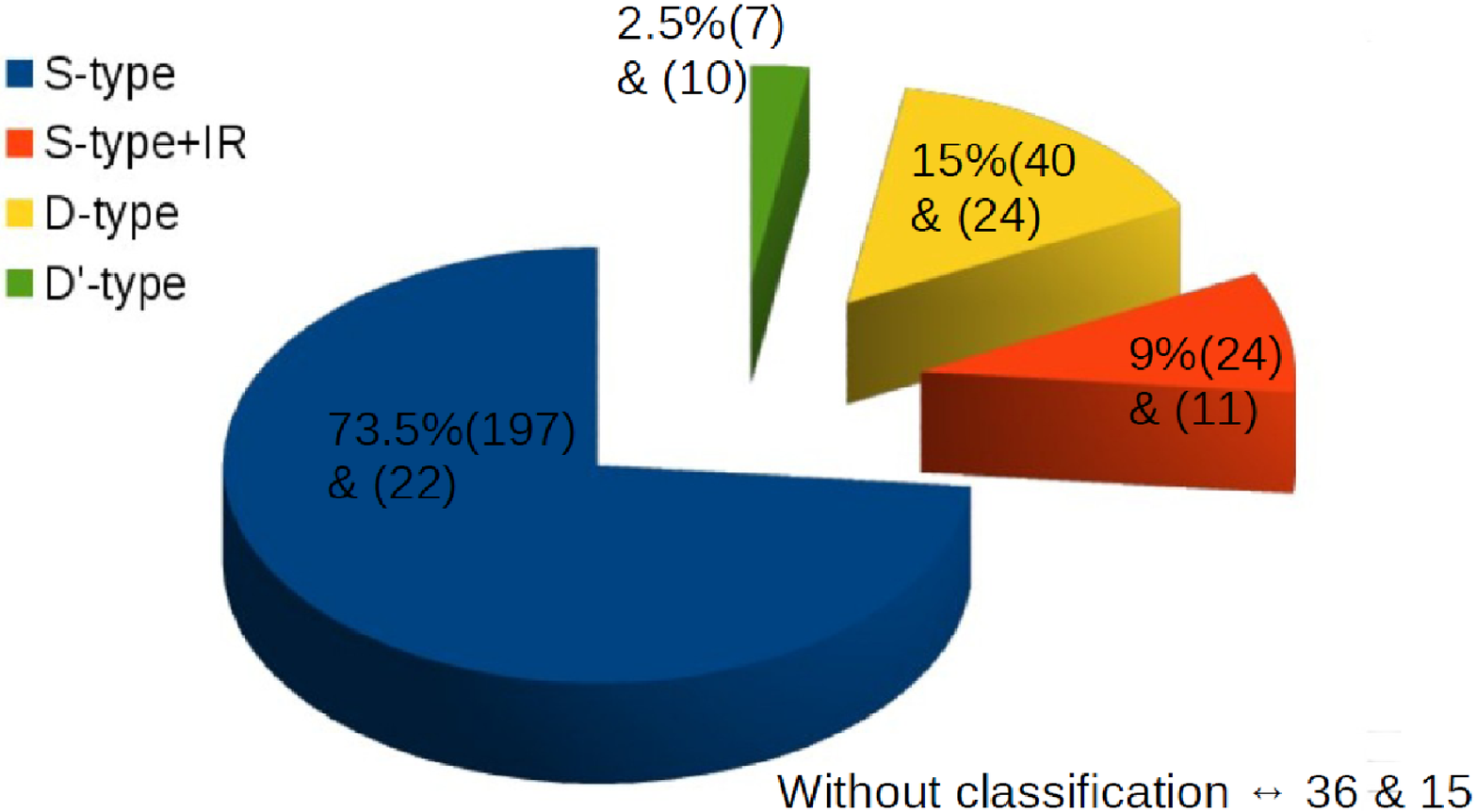}
\caption{Pie chart of the new classification of SySts. Numbers in parenthesis give the exact population of known and candidate SySts in each type. The 12 new Systs discoveries in M33 are not included. }
\label{}
\end{figure}

\vskip10pt
{\bf\large Acknowledgements}
I would like to thank the organizing committee for the opportunity to attend this event and present this work as 
well as for their financial support. S.A. and M.L.L.-F. also acknowledge support of CNPq, Conselho Nacional de 
Desenvolvimento Cient\'ifico e Tecnol\'ogico - Brazil (grant 300336/2016-0 and 248503/2013-8 respectively).

\end{document}